\documentclass[%
superscriptaddress,
 amsmath,amssymb,
 aps,
 prx,
 twocolumn,
 10pt
]{revtex4-2}

\usepackage{graphicx}
\usepackage{dcolumn}
\usepackage{bm}
\usepackage{comment}
\begin{document}


\title{Direct measurement of coherent nodal and antinodal dynamics in underdoped Bi-2212}%

\author{Rishabh Mishra}
\affiliation{Optical Sciences Centre, Swinburne University of Technology, Hawthorn, Australia}%
\affiliation{University of New South Wales, Sydney, Australia}
\author{Jonathan O. Tollerud}%
\affiliation{Optical Sciences Centre, Swinburne University of Technology, Hawthorn, Australia}%

\author{Paolo Franceschini}%
    \affiliation{Department of Information Engineering, University of Brescia, Italy
    }%
    
\author{Nikolas Stavrias}%
\affiliation{Optical Sciences Centre, Swinburne University of Technology, Hawthorn, Australia}%
\author{Fabio Boschini
}%
     \affiliation{ Quantum Matter Institute, University of British Columbia, Vancouver, Canada
    }%
     \affiliation{ Institut National de la Recherche Scientifique, Varennes, Quebec J3X 1S2, Canada
    }
\author{Genda Gu
}%
     \affiliation{ Materials Science Division of Brookhaven National Laboratory, Upton, NY, USA
    }%
\author{Andrea Damascelli
}%
     \affiliation{ Quantum Matter Institute, University of British Columbia, Vancouver, Canada
    }%
         \affiliation{Department of Physics \& Astronomy, University of British Columbia, Vancouver, Canada
    }%
    
\author{Daniele Fausti}%
    \affiliation{Department of Physics, Friedrich-Alexander-Universit$\ddot{a}$t, Erlangen, Germany
    }%
    
\author{Jared H. Cole}%
    \affiliation{School of Science, RMIT University, Melbourne, Australia
    }%
    
\author{Claudio Giannetti}%
    \affiliation{Department of Mathematics and Physics, Universit\`a Cattolica del Sacro Cuore, Brescia, Italy
    }%

\author{Jeffrey A. Davis}%
    \email{jdavis@swin.edu.au}
    \affiliation{Optical Sciences Centre, Swinburne University of Technology, Hawthorn, Australia}%

\date{\today}

\begin{abstract}
The physics of strongly correlated materials is deeply rooted in electron interactions and their coupling to low-energy excitations. Unraveling the competing and cooperative nature of these interactions is crucial for connecting microscopic mechanisms to the emergence of exotic macroscopic behavior, such as high-temperature superconductivity. 
Here we show that polarization-resolved multidimensional coherent spectroscopy (MDCS) is able to selectively drive and measure coherent Raman responses weighted toward different parts of the Fermi surface, where the superconducting gap vanishes or is the largest (respectively called Nodal and Antinodal region) in underdoped Bi-2212.
Our measurements reveal that in the superconducting phase, the energy of Raman response weighted towards the nodal region is anti-correlated with the energy of electronic excitations at $\sim$1.6~eV, 
and both maintain coherence for $>44\pm8$~fs. In contrast, Raman response weighted towards the antinodal region show significantly faster decoherence, limited by the experimental resolution ($<$20~fs), and no measurable correlations. 
Importantly, this long-lived coherence in the nodal region is specific to the superconducting phase and vanishes in the pseudogap and normal phases. 
This anticorrelation reveals a coherent link between the response at optical frequencies associated with transitions among hybridised O-2$p$ and Cu-3$d_{x^2-y^2}$ states
and low-energy excitations in the nodal region, sampled by the B$_{2g}$ Raman response. 
The different coherent dynamics for the B$_{1g}$ and B$_{2g}$ response in the superconducting phase are consistent with weaker dephasing of near-nodal excitations by strongly momentum-dependent scattering processes, including scattering from antiferromagnetic fluctutations. 

\end{abstract}

\maketitle

\section{\label{sec:level1}Introduction}

Unconventional superconductivity in copper oxides is arguably the major unsolved problem in condensed matter physics. 
The exact mechanism responsible for Cooper pair formation, and the role of different collective excitations remain unresolved. 
The complexity arises from electron-electron interactions and electronic interactions with phonons, magnons, and other collective fluctuations. These effectively change the local potential and balance of electron-electron interactions, leading to different types of macroscopic order, from superconducting Cooper pairs to charge ordered phases, antiferromagnetic Mott insulators, and more~\cite{keimer2015quantum, fradkin2015colloquium,lee2006doping,le2006two}. 

Adding to the challenge, several of these bosonic excitations, with typical energy scale of the order of meV, are intricately intertwined with excitations at much higher energy, up to 1-3~eV. The first indication of this was observed through the redistribution of spectral weight from high to low energy states, as the doping level is increased \cite{holcomb1994optical,holcomb1996thermal,molegraaf2002superconductivity,norman2003electronic, boris2004plane,basov2005electrodynamics}. Observations of an anti-correlation between the charge-transfer gap and superconducting transition temperature further pointed to a role for the high-energy states in impacting the formation of the superconducting (SC) phase \cite{ruan2016relationship,wang2023correlating}. Transient reflectivity measurements have provided additional understanding of these links. For example, excitation of Bogoliubov quasiparticles via Raman processes or directly by THz fields, leads to a change in the response at optical frequencies associated with transitions among the manifold of states 
that emerge from hybridization of O-2$p$ and Cu-$3d_{x^2-y^2}$ orbitals~\cite{Giusti2019,Giusti2021,Mansart2013,giannetti2011revealing,giannetti2016ultrafast,Madan2016,toda2014rotational}. Other transient reflectivity and angle-resolved photoemission spectroscopy (ARPES) measurements have shown selective coupling of specific phonons to these excitations in the 1-3~eV range \cite{Ishioka2023,Yang2019}. 

Interactions among the low-energy excitations are also expected to be important and have been observed through nonlinear THz spectroscopy \cite{katsumi2018higgs,chu2020phase,chu2023fano,Liu2024,Katsumi2024}. These have revealed, for example, dynamics of the Higgs mode, from which details of the interactions that modify the order-parameter can be ascertained, which contributes to understanding the microscopic mechanisms underlying Cooper pair formation\cite{katsumi2018higgs,chu2020phase}. 

One of the characteristics of cuprate superconductors is the \textit{d}-wave pairing symmetry and a gap that varies as a function of in-plane wavevector, reaching its maximum at the antinode ((k$_x$,k$_y)\sim(0,\pm\frac{\pi}{a})$ or $(\pm\frac{\pi}{a},0)$), and zero at the nodal points $\sim(\pm\frac{\pi}{2a}, \pm\frac{\pi}{2a})$~\cite{shen1993anomalously,damascelli2003angle,lin2006raising,hashimoto2014energy}. 
In the region around the nodes, the gap fills as the system is heated to the critical superconducting temperature (T$_c$), above which the superconductivity is quenched. Meanwhile, in the antinodal region an electronic gap remains open up to a higher temperature, T*. The phase arising between T$_c$ and T* is referred to as the pseudogap (PG) phase, and many questions around its origin and nature remain unresolved~\cite{madan2014separating, kanigel2008evidence, yang2008emergence, nakayama2009evolution, shi2009spectroscopic, kondo2011disentangling}. 
Time-resolved ARPES measurements have been able to measure dynamics at the nodal and antinodal region following photoexcitation. These have shown that the superconducting gap close to the nodes is more sensitive to photoexcitation and slower to recover, compared to the antinodal region~\cite{smallwood2014time,zonno2021time}. 
The role of macroscopic coherence, and the interactions that destroy it have also been inferred from these measurements. For example, the quasiparticle peak gets much stronger and sharper below T$_c$, which is attributed to enhanced quasiparticle coherence and related to the macroscopic coherence required for superconductivity\cite{Feng2000,damascelli2003angle}.

Selective excitation, and measurement of the momentum dependent coherent dynamics of quasiparticles, collective modes, and other low-energy excitations,  has the potential to shed further insight on the key interactions. 
Raman scattering with orthogonally polarized excitation and detection aligned to the Cu-Cu or Cu-O directions selectively probes the response with B$_{1g}$ and B$_{2g}$ symmetry. In cuprates these symmetries predominantly weight electronic states near the antinodal and nodal regions of the Brillouin zone, respectively, and therefore provide momentum-selective access to low-energy excitations in those regions \cite{le2006two,devereaux2007inelastic,loret2020universal}. 
Pump-probe transient-reflectivity, with the polarization of the measured signal set to be perpendicular to the incident probe beam, has been used to measure the dynamics of the nodal and antinodal response through the B$_{1g}$ and B$_{2g}$ symmetry selection \cite{PhysRevB.84.174516, toda2014rotational, madan2017dynamics, Giusti2019, Giusti2021}.
However, in those measurements with a linearly polarized pump, the dominant excitation is isotropic, with A$_{1g}$ symmetry, which can subsequently evolve to give a response with B (B$_{1g}$ or B$_{2g}$) symmetry \cite{toda2014rotational}. The dynamics that are subsequently measured are then significantly impacted by the dynamics of the isotropic A$_{1g}$ excitations.

Multidimensional coherent spectroscopy (MDCS) is a technique that has been developed to measure ultrafast coherent dynamics and resolve interactions in complex systems across physics, chemistry and biology~\cite{Mukamel2000,hybl2001two,smallwood2018multidimensional, tollerud2017coherent}. It has been used, for example, to understand interactions between excitons, electrons, polarons, polaritons, phonons, and more in two-dimensional semiconductors~\cite{Hao2016,muir2022interactions,huang2023,Policht2021}, and to resolve energy-transfer pathways and mechanisms in the protein complexes involved in photosynthesis~\cite{Fuller2014,Zigmantas2024,Turner2012}.

Here, we use polarization-resolved MDCS to probe coherent Raman responses with B$_{2g}$ and B$_{1g}$ symmetry, which predominantly sample the nodal and antinodal regions, respectively, 
in underdoped (T$_c$=~82~K) Bi$_2$Sr$_2$CaCu$_2$O$_{8+\delta}$ (Bi-2212) in the SC, PG and normal phase. 
Importantly, with this approach, we are able to selectively isolate Raman coherences excited and detected with B$_{1g}$ or B$_{2g}$ symmetry. This allows the measured dynamics to be directly associated with excitations in the antinodal and nodal regions of the Brillouin zone, respectively, while minimising contributions from the dominant A$_{1g}$ response. 
When cooling below T$_c$ we see a strong anti-correlation between the energy of low-energy Raman excitations in the nodal region and the energy of electronic excitations at $\sim$1.6~eV, with a long coherence time exceeding 44~fs. On the contrary, for Raman excitations at the antinode, or temperatures above T$_c$, the strength of the correlation is significantly weaker, and the measured decoherence is much faster, limited by the experimental resolution ($<20$~fs). 

These observations of a distinct difference in the coherent dynamics and correlations at the nodal region in the superconducting phase provide further insight into the role of decoherence, and the impact this may have on maintaining a coherent superconducting condensate.

\section{Experimental methods}

\begin{figure*}
\includegraphics[width=1\textwidth]{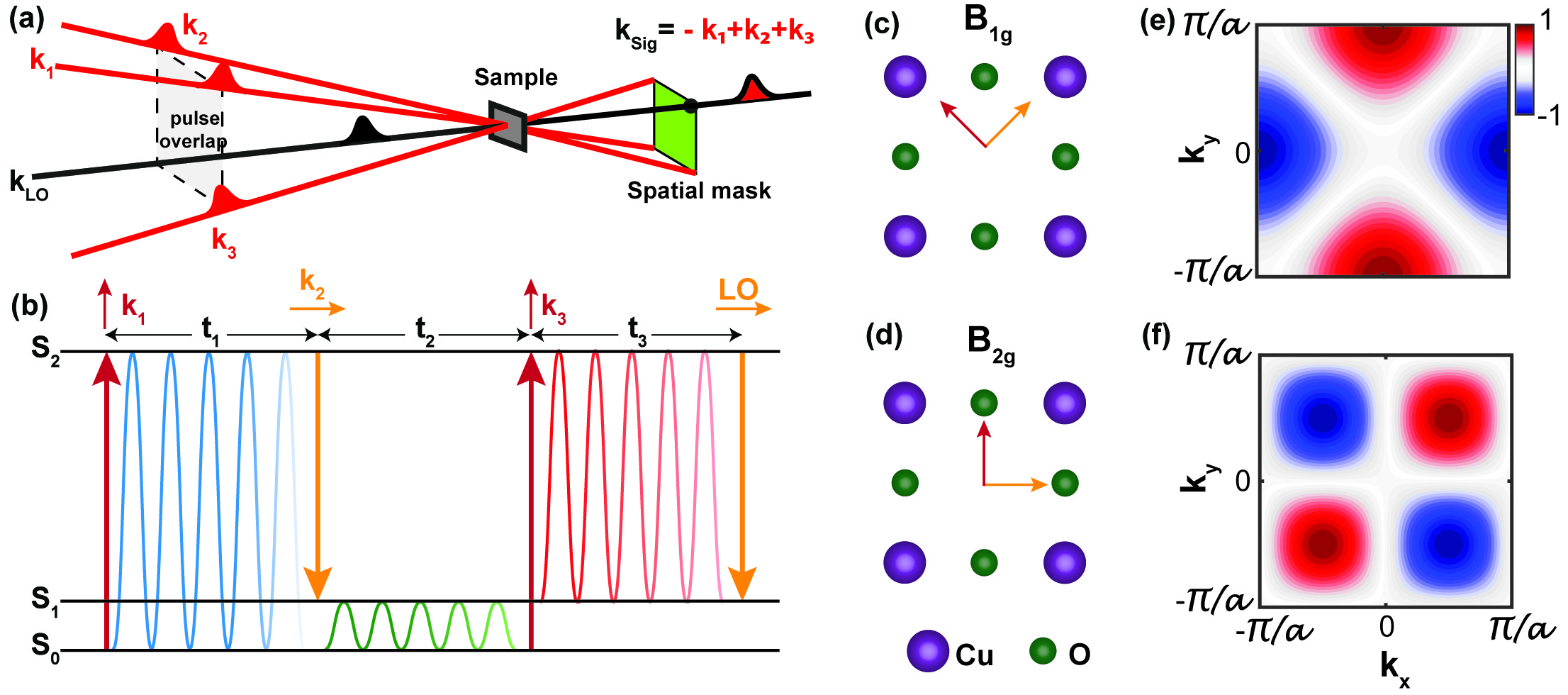}
\caption{ \textbf{The experimental configuration.} (a) the three excitation beams are arranged in a box geometry, meaning each beam is incident on the sample with a different wavevector (labeled \textbf{k$_{1-3}$}) with the signal being generated in the phase-matched direction, \textbf{-k$_1$ + k$_2$ + k$_3$}, appearing on the fourth corner of the box and overlapped with the local oscillator (\textbf{k$_\mathrm{{LO}}$}) to enable interferometric detection. The schematic shown here depicts the transmission geometry for clarity but the measurements were recorded in a reflection geometry, with a bulk sample. (b) The time-ordering of the pulses in the rephasing configuration used here, and the state of the system after each pulse in an isolated few-level system picture. The polarization of the \textbf{k$_1$} and \textbf{k$_3$} pulses are parallel to each other, but perpendicular to \textbf{k$_2$} and the LO, and are aligned with either (c) the Cu-Cu, or, (d) Cu-O, direction, leading to excitation and detection of the Raman response with (e) B$_{1g}$,  or (f) B$_{2g}$, symmetry, whose Raman vertices are concentrated near the antinodal and nodal regions of reciprocal space, respectively.}
\label{fig:0}
\end{figure*}

The MDCS technique is based on transient four-wave mixing and, similar to pump-probe transient absorption (TA) or transient reflectivity (TR) measurements, is dependent on the third-order susceptibility \cite{Mukamel2000,hybl2001two,smallwood2018multidimensional, tollerud2017coherent}. These measurements are typically analysed and interpreted in the context of isolated few-level systems, but in many-body systems with collective excitations, it can get more complicated \cite{zsbw-2yzb,Salvador2024,Liu2024}. However, to gain a conceptual understanding of these measurements we continue our description using the perturbative approach based on a few-level system. In this framework, for TA/TR measurements two light-matter interactions come from a single `pump' pulse. 
In MDCS, this is split into two separate interactions, which can arrive at different times (Fig.~\ref{fig:0}~(b)). 
The first pulse excites a coherent superposition between two electronic states which evolves as a function of time, $t_1$, until the second pulse arrives. 
The interaction with the second pulse converts this coherence into a population in the excited or ground state, or a Raman coherence in either the ground or excited state manifold. This, in turn, is transformed by the third pulse into a coherent superposition of two different electronic states that then radiates the third-order signal. The amplitude and phase of this third-order response are acquired by interfering the signal with a fourth (local oscillator) pulse, which is attenuated to be 3 orders of magnitude lower intensity than the excitation pulses, and measuring a spectral interferogram in a spectrometer. 
One of the advantages of MDCS measurements over TA/TR is that they provide direct access to coherent dynamics, which are more sensitive to interactions and can reveal details of many body effects and correlations between different energy and time scales.

In our MDCS measurements, we adopt a non-collinear geometry with each of the three excitation beams incident on the sample with different wavevectors, labelled \textbf{k$_1$}, \textbf{k$_2$}, \textbf{k$_3$}, as depicted in Fig.~\ref{fig:0}~(a). We then collect the signal emitted in the momentum-conserving direction, -\textbf{k$_1$}+\textbf{k$_2$}+\textbf{k$_3$}. 
A key advantage of the non-collinear approach is that it allows independent control of the polarization of each pulse, so, for example, the first two pulses can have different polarizations, allowing selective excitation of Raman coherences with B$_{1g}$ or B$_{2g}$ symmetry \footnote{More correctly, the measured signal is only sensitive to the initial excitation of coherent Raman response with B$_{1g}$ or B$_{2g}$ symmetry}.
Similarly, the polarization of the \textbf{k$_3$} pulse and the emitted signal can be chosen to selectively detect Raman responses with the same symmetry. (see Section-IX in Supplementary material for detailed calculation of the Raman selection rules for these MDCS measurements).

The arrival time of each pulse is individually controlled using a pulse shaper based 2D spatial light modulator \cite{tollerud2017coherent,tollerud2014isolating,Turner2011}. The pulse shaper is also used to compress each of the beams using the Multiphoton intrapulse interference phase scan (MIIPS) algorithm \cite{xu2006quantitative}. 
Measurements were recorded as a function of the delay between the first two pulses, $0\leq t_1\leq 200$~fs, the delay between the second and third pulse, $-50 \leq t_2\leq300$~fs, and the emission energy, $\hbar\omega_3$, which is directly resolved by the spectrometer that measures the spectral interferograms. Details of data processing steps to isolate the amplitude and phase of the third order signal, $E_{sig}(t_1,t_2,\hbar\omega_3)$ from the measured spectral interferograms are described in the Supplementary material (Section-III). A Fourier transform of $E_{sig}(t_1,t_2,\hbar\omega_3)$ with respect to $t_1$ yields a series of 1Q 2D spectra that correlate what is effectively the absorption energy ($\hbar\omega_1$) and the signal emission energy ($\hbar\omega_3$) for a given value of t$_2$. This type of 2D spectrum has been used extensively in semiconductors and molecular systems to reveal coupling, interactions, and charge/energy transfer between well-defined electronic states \cite{smallwood2018multidimensional,tollerud2017coherent,tollerud2016revealing,nardin2014coherent,huang2023,Policht2021,muir2022interactions}. 
An additional Fourier transform with respect to $t_2$ yields a 3D spectrum, which separates Raman coherences along the $\hbar\omega_2$ axis, and can reveal correlations of these Raman excitations with optical resonances evident in the $\hbar\omega_1$ and/or $\hbar\omega_3$ dependence~\cite{davis2011three,tollerud2014isolating,Kolesnichenko2019,Hao2016,Yang_2DRaman}.



For the measurements reported here, laser pulses centred at a wavelength of 770 nm (1.61 eV), 
with near transform-limited duration of $\sim$20 fs, were used to excite the sample and drive the nonlinear response (see Section-IV of the Supplementary Material for details of the pulse characterization). As depicted in Fig.~\ref{fig:0}(b-d), the \textbf{k$_1$} and \textbf{k$_3$} beams were polarized perpendicular to \textbf{k$_2$} and LO, and aligned with the Cu-Cu (Cu-O) direction to selectively excite Raman coherences with B$_{1g}$ (B$_{2g}$) symmetry.  
For the measurements reported in the SC and PG phase (at T= 40~K and 90~K, respectively), the fluence was kept below $\sim$4~$\mu$J\text{/}cm$^2$ per pulse to ensure a weak perturbative regime, i.e. where the excitation does not drive a photo-induced phase transition \cite{Giannetti2009,Coslovich2011,Kusar2008} (see Section-I in supplementary material for more details).

\begin{figure*}
\includegraphics[width=\textwidth]{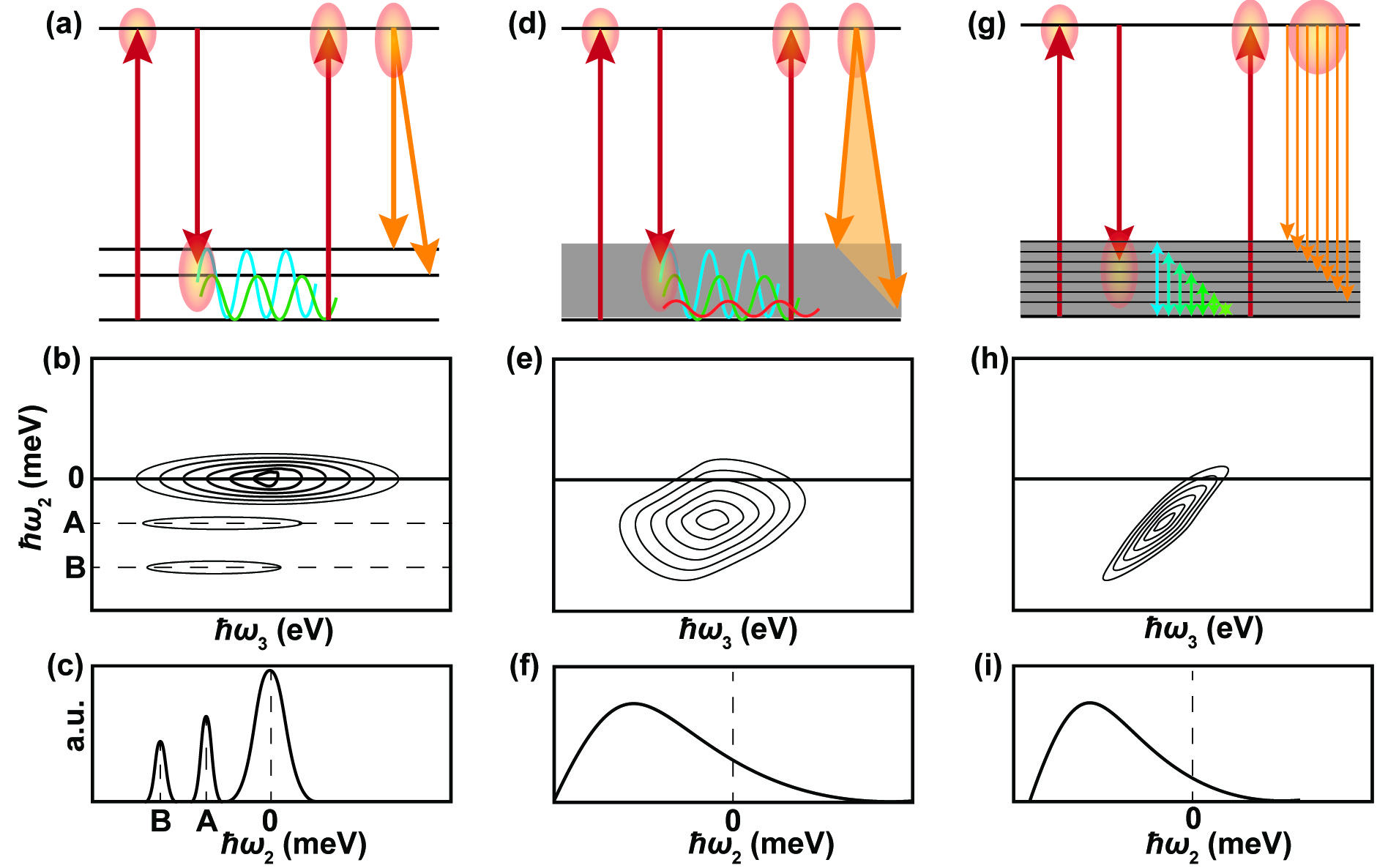}
\caption{\textbf{Interpreting 0Q 2D spectra for different Raman responses}. (a-c) shows the case where there are discrete vibrational modes, (d-f) for a broad uncorrelated Raman response, and (g-i) for a broad Raman response where the transitions at low and high energy are anti-correlated. The top row (a,d,g) shows the Stokes Raman pathways with the rephasing pulse ordering used here, with two discrete Raman modes, a broad Raman response, and an example of a broad Raman response with energies that are anti-correlated with transitions to the optically excited state. Each of these shows distinct features in the 0Q 2D spectra (middle row, (b,e,h)), but in the 1D Raman spectrum (c,f,i), there is no difference between the cases of correlated and uncorrelated broadening. 
}
\label{fig:1}
\end{figure*}

To visualise the Raman excitations and correlations with the electronic transitions, we plot the amplitude of the signal as a function of $\hbar\omega_2$ and $\hbar\omega_3$. 
In this type of 2D spectrum (referred to as a zero-quantum, or 0Q, spectrum), a system with discrete Raman modes, and in the absence of any resonant enhancement from allowed optical transitions, will produce narrow horizontally-shaped peaks \cite{Kolesnichenko2019} as depicted in Fig.~\ref{fig:1}~(a-c). Integrating the signal along the $\hbar\omega_3$-axis yields a response similar to typical coherent Raman measurements. The negative $\hbar\omega_2$ values indicate the signal is generated via coherent Stokes Raman processes.  
In contrast, a system with a broad Raman response, such as that seen in previous Raman measurements on Bi-2212~\cite{loret2020universal,yang2020ultrafast}, will normally give a broad peak in the 0Q MDCS spectrum, as depicted in Fig.~\ref{fig:1}~(d-f).
However, if there is some correlation of the energies of the Raman transition/s and the optical transitions, the MDCS spectrum can show a diagonally-shaped peak \cite{novelli2020persistent}, as illustrated in Fig.~\ref{fig:1}~(g-i) for anti-correlated transition energies. In this case, the narrow, cross-diagonal width can be limited by dephasing of either of the transitions or the strength of the correlation.
Notably, conventional Raman measurements cannot distinguish between correlated and uncorrelated contributions to a broad excitation continuum, as illustrated in Fig. 2~(f,~i). The emergence of a diagonal feature in the 0Q MDCS spectrum therefore provides direct evidence that the low- and high-energy excitations are not independent, but are linked through a well-defined correlation.

One way of generating a 0Q 2D spectrum is to set $t_1=0$ and scan $t_2$. However, this gives equal weighting to rephasing and non-rephasing pathways (i.e. where the \textbf{k$_1$} pulse interacts first or second, respectively) \cite{Mukamel2000,hybl2001two,tollerud2017coherent,Kolesnichenko2019}, making it more difficult to extract and isolate specific signal contributions. To reduce the number of pathways contributing to the signal, we instead acquire a 3D rephasing spectrum and integrate the response as a function of $\hbar\omega_1$ (see Section-II in supplementary materials for details). One of the outcomes is that for ground state Raman pathways, the Stokes Raman signal (with $\hbar\omega_2 < 0$) is preferred, as has been depicted in the schematics in Fig.~\ref{fig:1} and explained in Section-II of the Supplementary Material.

In these measurements, the range of $\hbar\omega_2$ and $\hbar\omega_3$ values that can be accessed is shaped by the laser pulse spectrum, and are not completely independent. The instrument window function (IWF), which defines the scaling of the signal due to the laser spectrum, can be calculated as described and plotted in Supplementary material section V. The IWF is used as a reference and compared to the measured response from the Bi-2212 sample.

\section{Results and discussion}

\begin{figure*}
\includegraphics[width=\textwidth]{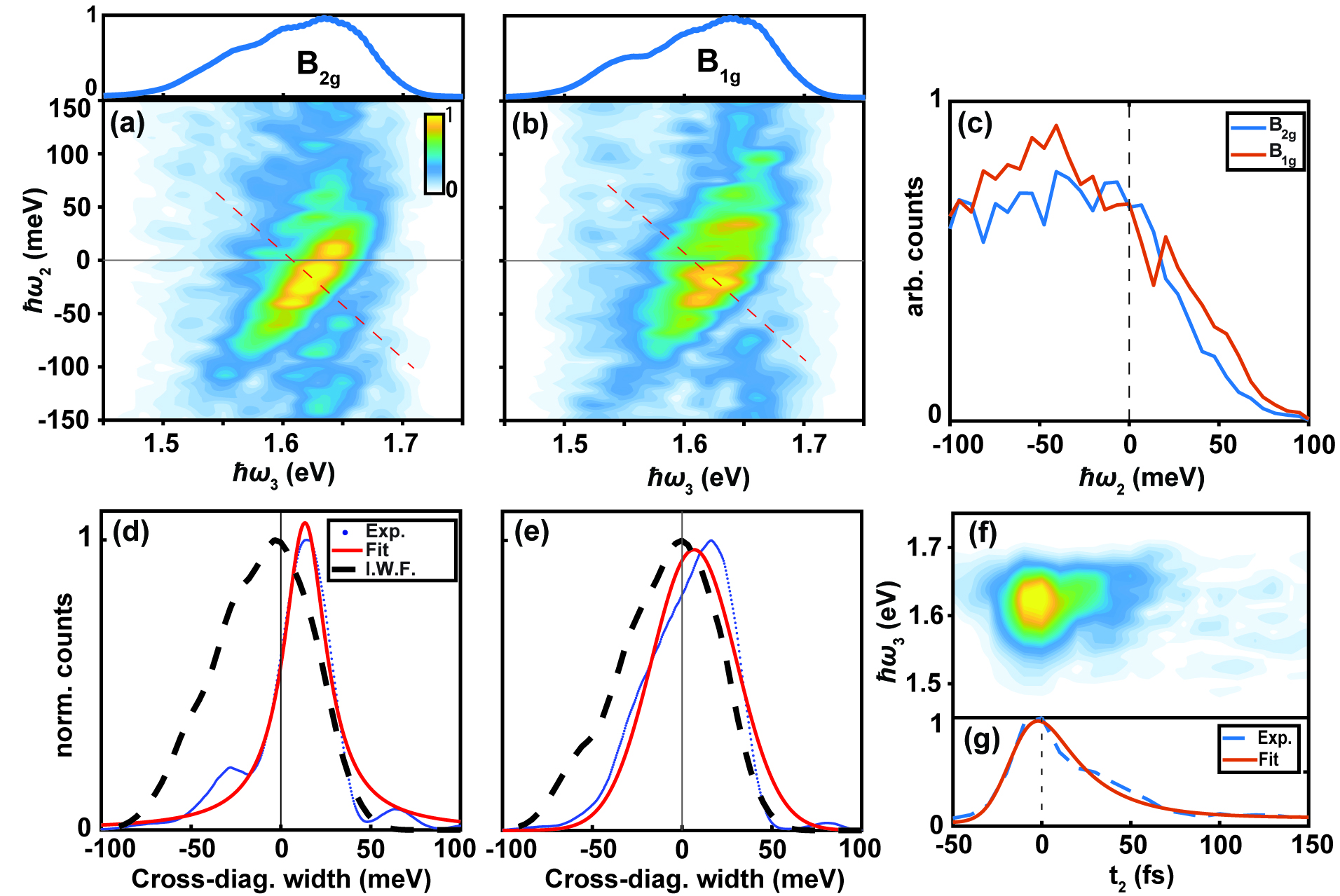}
\caption{ \textbf{Measurements obtained in the superconducting phase.} (a) and (b) show the 0Q 2D spectrum obtained by integrating the 3D rephasing spectrum with respect to $\hbar\omega_1$ for the B$_{2g}$ and B$_{1g}$ geometry, respectively. (c) Shows the measured Raman response obtained by integrating the 2D data with respect to $\hbar\omega_3$, and normalising with respect to the instrument window function (this is taken as the average from the multiple repeats of the measurements shown in the Supplementary Material). The narrow cross-diagonal lineshape in the B$_{2g}$ configuration is highlighted in (d) and (e), which shows the cross-diagonal slices (taken in the direction of the dashed lines in (a) and (b)) for the B$_{2g}$ and B$_{1g}$ configurations, respectively.  These are fit with a Lorentzian (red line) and compared to the instrument window function (IWF, dashed line).  The narrow cross diagonal linewidth of 30$\pm7$~meV in the B$_{2g}$ geometry, corresponds to a decay time constant of 44$\pm8$~fs, and maps on to dynamics observed in the time domain, obtained by Fourier transforming the data with respect to $\hbar\omega_2$, shown as a function of $\hbar\omega_3$ in (f) and integrated in (g) }
\label{fig:2}
\end{figure*}

Figure~\ref{fig:2}~(a, b) show the 0Q-2D spectra in the SC phase (T=~40~K) for the B$_{2g}$ and B$_{1g}$ configurations, respectively. In both cases, a broad Raman response is measured, typical of Bi-2212, with the signal extending from $\hbar\omega_2\sim40$ to beyond $-80$~meV, limited on the negative side by the bandwidth of our laser pulses (see Section-VII in supplementary material for details of the accessible range of energies and the instrument window function). It is evident that the majority of the signal appears at $\hbar\omega_2<0$, which confirms that the dominant signal pathway is the ground-state Stokes Raman coherence pathway.  This is also evident in the 1Q 2D spectrum in the Supplementary Material (see Section-V for details), which shows the majority of the signal arises for $|\hbar\omega_1|>\hbar\omega_3$.

To more closely examine and consider the origin of this Raman response, the signal is integrated over $\hbar\omega_3$ and normalised by the instrument window function, as shown in Fig.~\ref{fig:2}~(c). These plots show a broad response as a function of $\hbar\omega_2$, corresponding to the B$_{1g}$ and B$_{2g}$ Raman spectra. 
Previous Raman measurements on Bi-2212~\cite{loret2020universal} have shown that this broad continuum contains contributions from an electronic continuum arising from particle-hole excitations in the correlated electronic system, shaped by interaction and self-energy effects, with the B$_{1g}$ and B$_{2g}$ symmetry responses accessing distinct momentum-space projections of that continuum. Below T$_c$, 
superconductivity-induced features have been shown to appear on top of the electronic continuum. The B$_{1g}$ response (at $\sim$70~meV) predominantly samples the larger gap scale near the antinode, whereas the B$_{2g}$ response (at $\sim$50~meV) samples an effective superconducting energy scale from the near-nodal sector.
~\cite{loret2020universal}. 
In our measurements, we cannot distinguish a discrete pair-breaking peak in either the B$_{1g}$ or B$_{2g}$ Raman response, however, it is likely that part of the observed response is associated with pair-breaking transitions and the resulting coherence between Cooper pairs and Bogolioubov quasiparticles\cite{Anderson1958,Mansart2013}. 

While the broad Raman response along $\hbar\omega_2$ is quite similar for the two configurations, the shape of the peaks in the 2D spectra are different. 
The response with B$_{2g}$ symmetry shows a narrow lineshape along the anti-diagonal direction, indicative of energy-energy anti-correlation between low-energy excitations in the nodal region and optical excitations at $\sim1.55-1.65$~eV. Conversely, in the B$_{1g}$ configuration the response is much broader, and although it still has a diagonal shape, this more closely matches the instrument window function (Fig.~S6 in supplementary materials). The cross-diagonal width can be quantified by taking slices perpendicular to the diagonal direction (i.e. along the direction of the dashed lines in Fig.~\ref{fig:2}~(a, b)). The average of 10 slices taken at different points across the peak are plotted in Fig.~\ref{fig:2}~(d, e) along with a Lorentzian fit to the data. The equivalent profiles for the instrument window function, determined on the basis of the laser spectrum used for each measurement, are also shown for comparison. These measurements were repeated on multiple occasions on different regions on the sample, and in each case a narrow diagonal lineshape was observed in the B$_{2g}$ configuration, and a broader profile similar to the instrument window function was seen for the B$_{1g}$ configuration (see Section-VIII in supplementary material).  The cross-diagonal linewidth (full width at half maximum) determined for the B$_{2g}$ configuration was $30 \pm 7$~meV while for the B$_{1g}$ configuration it was $70 \pm 10$~meV. The narrow cross-diagonal width for the B$_{2g}$ configuration is also reflected in the signal dynamics as a function of $t_2$, as shown in Fig.~\ref{fig:2}~(f, g). The cross-diagonal width and fit of the dynamics give a decay constant of 44$ \pm 8$~fs for the B$_{2g}$ configuration, and an upper limit of $20\pm 3$~fs for B$_{1g}$, limited by the pulse duration.

Significantly, the narrow width and $>44\pm8$~fs decoherence time are only measurable because of the anti-correlation between $\hbar\omega_2$ and $\hbar\omega_3$. 
In 1Q 2D spectra, narrow diagonal peaks typically arise due to rephasing (i.e. an unwinding of the effects of inhomogeneous broadening/disorder) \cite{tollerud2017coherent,muir2022interactions,smallwood2018multidimensional,nardin2014coherent,Mukamel2000,hybl2001two,Liu2024}. In isolated few-level systems this allows separation of homogeneous and inhomogeneous broadening. \footnote{In rephasing measurements of collective excitations, the signal pathways are more complicated, and the 2D diagonal peak shape can instead be used to separate elastic and inelastic scattering mechanisms \cite{zsbw-2yzb,Salvador2024,Liu2024}}. In contrast, for the 0Q 2D spectra presented here, the diagonal peak shape arises as a result of correlations, and decoherence times longer than the pulse durations. As discussed above, the cross-diagonal width can be limited by one or more of: dephasing of the Raman ($\hbar\omega_2$) coherence, dephasing of the electronic ($\hbar\omega_3$) coherence, or the strength of the correlation between the excitations at the two energy scales. 
The observation here of a diagonal peakshape only in the B$_{2g}$ configuration identifies a coherent part of the broad low-energy response in the nodal region that is coherently linked to optical transitions at $\sim$1.6eV, and which cannot be resolved with linear Raman spectroscopy. 

\begin{figure*}
\includegraphics[width=\textwidth]{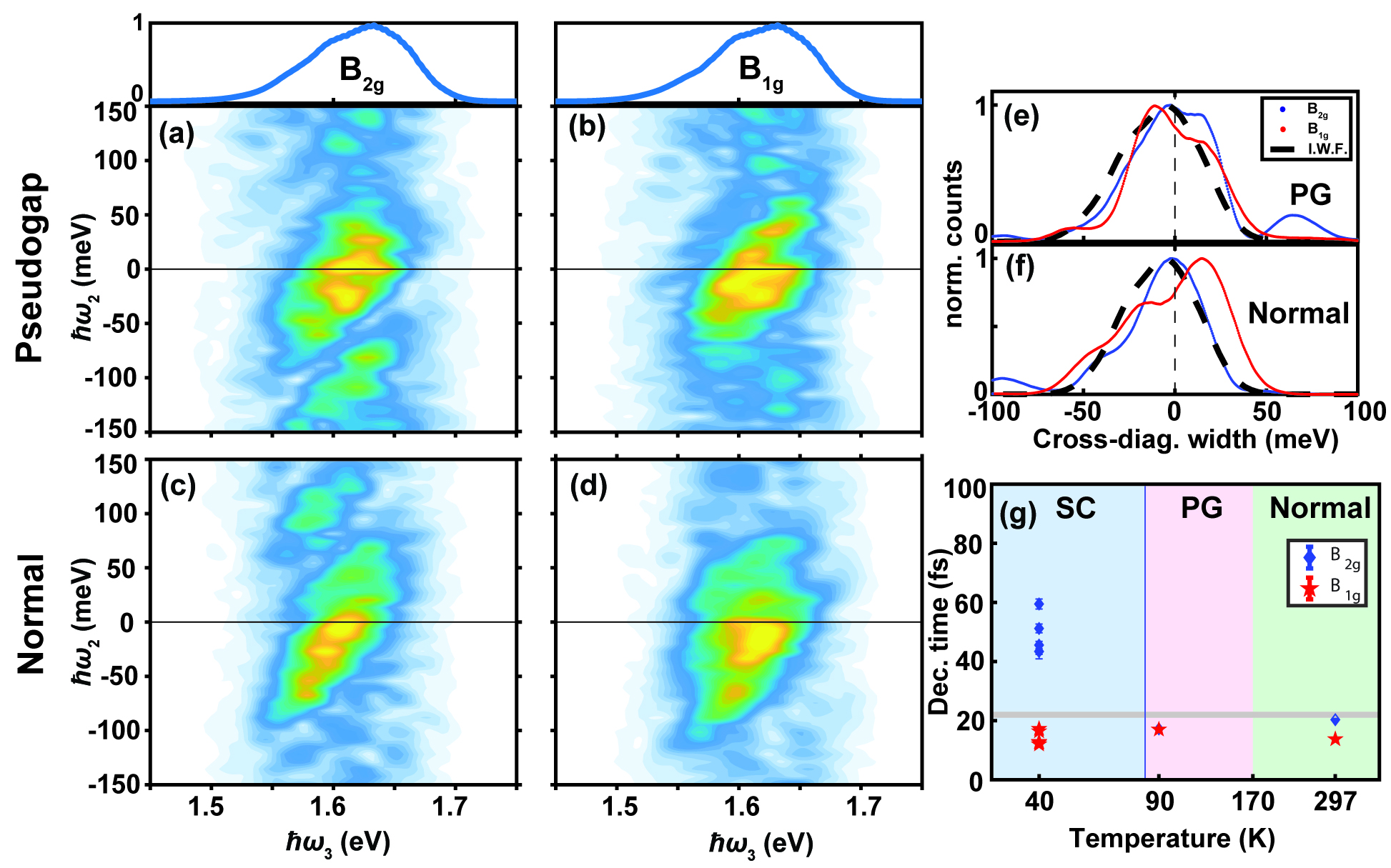}
\caption{\textbf{Measurements obtained in the pseudogap and normal phases.}(a) and (b) show the 0Q 2D spectra at T=~90~K, in the PG phase for B$_{2g}$ and B$_{1g}$ configurations, respectively. (c) and (d) show the same at T=~290~K, in the normal phase. (e) and (f) show cross-diagonal slices for the B$_{2g}$ (blue) and  B$_{1g}$ (green) data in the PG and normal phase, respectively, along with the slice from the instrument window function (dashed line). These show that the widths closely match the instrument window function. (g) summarises the decoherence times determined from the cross-diagonal slices for the different symmetries, temperatures and repeated measurements.}
\label{fig:3}
\end{figure*}

To help understand the origin of the correlations and enhanced coherence in the nodal region, these measurements were repeated at different temperatures to compare the dynamics in the PG (T=~90~K) and normal (T=~293~K) phases. In both cases, shown in Fig.~\ref{fig:3}, there is no indication of a narrow diagonal peak in either the B$_{1g}$ or B$_{2g}$ configuration and the response closely matches the IWF. 

It is clear from these measurements that the narrow diagonal peak is only observed in the SC phase and when probing the low-energy Raman response with B$_{2g}$ symmetry (Fig.~\ref{fig:2}(a)). When moving to the antinodal region or higher temperatures (where the sample is in the pseudogap or normal phase) the decoherence rate increases by more than a factor of 2, and/or there is a much reduced correlation between the Raman and optical excitations.
Because the decoherence and energy correlations are convolved in the 0Q 2D spectra, it is difficult to separate the changes to these two different effects.
To understand the hierarchy of these processes and how they change with the phase of the system, we consider the nature and interactions of the Raman excitations, the optical transitions, and the possible origins of the correlations. 

As discussed above, the Raman response is expected to involve contributions from particle-hole excitations in the shaped by interaction and self-energy effects, 
as well as some contribution from pair-breaking excitations and the associated coherence between Cooper pairs and Bogoliubov quasiparticles.  
The present MDCS measurements isolate the coherent component of the low-energy response sampled by the B$_{1g}$ and B$_{2g}$ Raman tensor.  
This is achieved because the MDCS protocol used here only measures the Raman response with B$_{2g}$ or B$_{1g}$ symmetry that arises from coherent Raman excitations of the same symmetry. 
In contrast, previous approaches \cite{PhysRevB.84.174516, toda2014rotational, madan2017dynamics, Giusti2019, Giusti2021} that have measured an anisotropic B$_{2g}$ or B$_{1g}$ response have done so following an initial isotropic A$_{1g}$ excitation and subsequent anisotropic evolution \cite{toda2014rotational}. 
To highlight the contrast, measurements performed with co-linearly polarised pulses are shown in the Supplementary Material  (Fig.~S5). These are dominated by the A$_{1g}$ Raman response and yield a signal more than two orders of magnitude stronger than those obtained with cross-linearly polarised excitation.
With the greater signal selectivity in the measurements presented in Fig.~\ref{fig:2} and Fig.~\ref{fig:3}, we are able to directly isolate the coherent dynamics associated with the chosen symmetry.

The enhanced coherence observed in the response with B$_{2g}$ symmetry is consistent with other measurements that have suggested increased coherence of nodal quasiparticles below T$_c$ \cite{Giusti2019,Giusti2021,yamasaki_unmasking_2007,dahm_nodal_2005}. 
This has been attributed to antiferromagnetic fluctuations preferentially scattering antinodal rather than nodal quasiparticles~\cite{dahm_nodal_2005,cilento_dynamics_2018,zhu2022spin}, resulting in significantly stronger decoherence in the antinodal regions. It is reasonable to suggest that the narrow diagonal feature reflects the longer coherence times of excitations within the nodal region of momentum space, however, we are only able to resolve these slowed coherent dynamics because of the anti-correlation with the optical transition energy.

Strong links between excitations across the different energy scales have been shown previously, and are an accepted, though much debated, aspect of superconductivity in cuprates. 
This arises primarily from coupling of the O-2\textit{p} and Cu-3\textit{d$_{x^2-y^2}$} orbitals, which leads to the formation of 
a manifold of hybridized Cu-O states. 
In linear spectroscopy measurements, this is seen as a redistribution of the spectral weight as the doping is increased, from the charge transfer (CT) peak at 1-3~eV to a continuum of lower-energy transitions~\cite{molegraaf2002superconductivity,boris2004plane,basov2005electrodynamics}. In TA/TR measurements, it has been shown that melting of the superconducting phase (either thermally, or as a result of photoexcitation) leads to a significant change in the transient reflectivity at energies close to and below the CT gap 
\cite{gedik_abrupt_2005,giannetti2011revealing,giannetti2016ultrafast}.

On the basis of transient reflectivity measurements in lanthanum strontium copper oxide (LSCO), Mansart et al. also made a direct link between pair-breaking excitations in the nodal region at $\sim$20~meV and the charge-transfer transition at 2.6~eV \cite{Mansart2013}. This was based on oscillations observed in the transient response that they attributed to Raman 
driven coherent oscillations of the Cooper-pair condensate. The frequency of these oscillations closely matched the expected energy for the pair-breaking transition, and when the polarization was such that the A$_{1g}$ and B$_{2g}$ response was selected, they identified a strong enhancement of the oscillations for a probe energy of 2.6~eV, consistent with the CT transition in LSCO.  No such enhancement was seen for polarizations selecting the B$_{1g}$ symmetry excitations, suggesting that only in the nodal region are the excitations at 2.6~eV and $\sim$20~meV linked.  It is worth noting, however, that in those measurements the B$_{2g}$ excitations cannot be isolated from excitations with A$_{1g}$ symmetry, and thus contributions from those A$_{1g}$ excitations cannot be excluded. Additionally, the fluence of the pump (300~$\mu$J\text{/}cm$^2$) was far in excess of the threshold for melting the superconducting phase \cite{Giannetti2009,Coslovich2011,Kusar2008}, which significantly complicates the picture. 

In the measurements reported here, the cross-polarized \textbf{k$_1$} and \textbf{k$_2$} beams ensure that only Raman coherences with B$_{2g}$ symmetry (or B$_{1g}$ in the other configuration) can contribute to the signal, and the fluence used (4~$\mu$J\text{/}cm$^2$) was kept well below the photo-melting threshold. 
The observed spectral anti-correlation demonstrates that the energy of low-energy excitations contributing to the coherent B$_{2g}$ Raman response is linked to the energy of optical transitions in the 1.55-1.65 eV range. The microscopic origin of this relationship remains to be established, but it implies that the low-energy nodal excitations and the hybridised Cu-O electronic structure are not independent degrees of freedom.

From this perspective, the anti-correlation observed in the 0Q spectra provides more than evidence of coupling between low- and high-energy excitations. It demonstrates that variations in the energy of the low-energy nodal excitations are directly linked to variations in the energy of optical transitions within the hybridised Cu-O electronic structure. Such behaviour would not be expected if the low-energy and optical excitations evolved independently. Instead, the observed link between the two energy scales points to a common many-body origin and highlights the importance of the hybridised Cu-O electronic structure in determining the low-energy excitation spectrum associated with superconductivity.

\section{Conclusions}

The polarization-resolved MDCS measurements reported here have been able to reveal reduced decoherence and an energy-energy anti-correlation between the Raman response in the $\leq$~80~meV range and optical transitions at $\sim$~1.6~eV, which are present only in the superconducting phase and when the Raman response with B$_{2g}$ symmetry is selectively and coherently excited and probed.  
The broad Raman continuum of Bi-2212 is generally understood to contain contributions from particle-hole excitations in the correlated electronic system, shaped by interaction and self-energy effects. 
The present MDCS measurements indicate that a subset of these low-energy excitations, weighted toward the nodal region by the B$_{2g}$ Raman response, exhibit a well-defined anti-correlation with the energy of optical transitions near 1.6 eV. They thus reveal a coherent component of the low-energy excitation spectrum that is largely hidden within the broad Raman continuum measured using conventional linear spectroscopy. 

With the present measurements we do not identify the specific microscopic excitation responsible for the low-energy Raman response, or the anti-correlated transition around 1.6~eV. However, these measurements establish the existence of a coherent relationship between low-energy nodal excitations and optical transitions within the manifold of hybridised Cu-O states,
and provide a benchmark for theories describing the interactions between many-body states and collective excitations. 

Future measurements with finer temperature resolution across T$_c$ and as a function of doping will allow further comparison of the role of correlations and decoherence across the phase diagram and provide additional constraints for theory. Further questions around whether decoherence or correlations limit the cross-diagonal linewidth (and whether the two are linked) may be addressed by higher-order experiments capable of inducing rephasing of the Raman coherence~\cite{harel_four-dimensional_2017}.
Finally, and beyond the specific observations reported here, we note that these measurements demonstrate how MDCS with near infra-red (NIR) pulses can provide insight to help disentangle interactions in cuprate superconductors, and point to the potential of these approaches to enable greater understanding of strongly correlated electron materials.



\bibliography{Bi2212Coherence}

\bigskip
\begin{acknowledgments}
This work was funded by the Australian Research Council Discovery Project (DP210102050). The work at BNL was supported by the US Department of Energy,
office of Basic Energy Sciences, contract no. DOE-SC0012704.
This research was undertaken thanks in part to funding from the Max Planck–UBC–UTokyo Centre for Quantum Materials and the Canada First Research Excellence Fund, Quantum Materials and Future Technologies. This project is also funded by the Natural Sciences and Engineering Research Council of Canada (NSERC), the Canada Foundation for Innovation (CFI); the British Columbia Knowledge Development Fund (BCKDF); the Department of National Defence (DND); the Canada Research Chairs Program (F.B., A.D.), and the CIFAR Quantum Materials Program (A.D.).
\end{acknowledgments}

\end{document}